\documentclass[conference]{IEEEtran}
\IEEEoverridecommandlockouts
\usepackage{cite}
\usepackage{amsmath,amssymb,amsfonts}
\usepackage{bbm}
\usepackage{algorithmic}
\usepackage{graphicx}
\usepackage{textcomp}
\usepackage{eurosym}
\usepackage{multirow}
\usepackage{subcaption} 
\usepackage{xcolor}
\def\BibTeX{{\rm B\kern-.05em{\sc i\kern-.025em b}\kern-.08em
    T\kern-.1667em\lower.7ex\hbox{E}\kern-.125emX}}
\DeclareUnicodeCharacter{2212}{-}

\usepackage[draft=False]{hyperref} % put False when you've finished all your writing
\hypersetup{
     colorlinks=true,
     linkcolor=blue,
     filecolor=blue,
     citecolor=black,      
     urlcolor=cyan,
     }

\begin{document}

\title{Analysis of the French system imbalance paving the way for a novel operating reserve sizing approach}

\IEEEoverridecommandlockouts
\IEEEpubid{\makebox[\columnwidth]{979-8-3315-1278-1/25/\$31.00~\copyright2025 IEEE  \hfill} \hspace{\columnsep}\makebox[\columnwidth]{ }}

% https://www.lut.fi/en/eem23/call-papers
% 
\author{\IEEEauthorblockN{Jonathan Dumas, Sébastien Finet, Nathalie Grisey, Ibtissam Hamdane, Paul Plessiez}
\IEEEauthorblockA{RTE R\&D, Immeuble Window 7C, Place du Dôme 92073 LA DEFENSE Cedex, France\\
\{jonathan.dumas, paul.plessiez, ibtissam.hamdane\}@rte-france.com}
}

\maketitle

\begin{abstract}
%One of the primary responsibilities of Transmission System Operators (TSOs) is to ensure system adequacy at all times. Since most operating reserves cannot be activated immediately, and some reserves must be sized and contracted in advance, predicting future imbalances is crucial for maintaining the security of supply. However, the ongoing energy transition is causing significant changes in the electricity system, making it increasingly unstable and unpredictable.

This paper examines the relationship between system imbalance and several explanatory variables within the French electricity system. The factors considered include lagged imbalance values, observations of renewable energy sources (RES) generation and consumption, and forecasts for RES generation and consumption. The study analyzes the distribution of system imbalance in relation to these variables. Additionally, an HGBR machine-learning model is employed to assess the predictability of imbalances and the explanatory power of the input variables studied.
The results indicate no clear correlation between RES generation or consumption and the observed imbalances. However, it is possible to predict the imbalance adequately using forecasts available a few hours before real-time, along with the lagged values of the imbalance. Predicting the imbalance a day in advance proves to be complex with the variables examined; however, the extreme quantiles of the imbalance used for reserve sizing and contracting can be predicted with sufficient accuracy. 
%Future research could involve incorporating variables that reflect the behavior of balancing responsible parties, such as market prices, and further refining the machine learning prediction model.
\end{abstract}

\begin{IEEEkeywords}
Security of supply; balancing; system imbalance; operating reserve sizing 
\end{IEEEkeywords}

\section{Introduction}

% Intro générale
Transmission System Operators (TSOs) continuously balance electricity generation and demand. To achieve this aim close to real-time, they use operating reserves defined under the current European legislation, the System Operation Guidelines, to cope with unexpected variations in demand and generation.  
Operating reserves are generating capacities (conventional power plants, hydro reservoirs, storage and batteries, renewable generation) and flexible consumption that can be activated upward or downward at the TSO’s request to face a system imbalance. 

% Focus sur la mFRR
The French TSO, RTE, determines the need for \textit{manual Frequency Restoration Reserve} (mFRR) based on a margin strategy \cite{eem2023dumas} that dynamically estimates two key values: the available upward and downward flexibility capacities in the electricity system, and the total flexible power required to ensure the security of supply. The procurement strategy for the necessary operating reserve combines contracted power amounts—sized to address significant shortages, such as the failure of the largest nuclear power plant (1500 MW)—with continuous estimates of available flexible capacities. These estimates are based on forecasts and operational production plans directly provided by power producers.

% On explique pourquoi la méthode de dim de la mFRR doit être revue
The growing share of renewable energy sources, such as wind and photovoltaic (PV) systems, makes it increasingly difficult to balance the electrical grid. This challenge is especially prominent when excess production is caused by low consumption levels alongside high renewable generation. Traditionally, electrical systems were designed to address production shortages using flexible conventional power plants. As a result, the current approach to sizing these systems needs to be reevaluated to accommodate this significant transition to renewable energy effectively.

% On explique le focus de cette étude : analyse des éventuels facteurs explicatifs du déséquilibre du système
This study aims to analyze the potential factors contributing to imbalances in the French system. This analysis will lay the groundwork for future research on a new method for sizing mFRR. It is essential for the sizing approach, which is based on margins, to take these factors into account to accurately predict future risks of imbalance and propose optimal sizing solutions.
The main contributions are threefold:
% Revoir les contributions de l'étude
1) We focus on the important issue of analyzing system imbalances about renewable energy and consumption forecasts, actual outcomes, and delays. Our goal is to identify potential correlations with observed system imbalances. This area remains underexplored in the literature, as only a few studies \cite{GOODARZI2019110827, DEVOS2019272, EICKE2021105455, kratochvil2016system} have addressed it;
2) A machine learning model predicts system imbalance using exogenous variables, allowing for the testing of various potential explanatory variables;
3) The combined approach of analyzing the statistical distributions of imbalance while utilizing a machine learning model for forecasting enables us to explore the potential correlations between explanatory and target variables. This method deepens our analysis, moving beyond just the mean of the distributions or simple correlation coefficients. Ultimately, it enhances our understanding of imbalance trends and helps identify directions for future research. 
%Preliminary results indicate that the French system imbalances correlate poorly with renewable and consumption forecast errors. However, predicting the extreme quantiles of imbalance (1\% and 99\% quantiles) is possible with relatively good accuracy in the day ahead. Finally, imbalance seems to depend on the year considered (day, week, month, and season), which indicates that operating reserve sizing should vary depending on the time of year.

The paper is organized as follows: Section \ref{pbst} outlines the problem and describes the data used for the analysis. Section \ref{sec:imb-analysis} provides an analysis of the imbalance. Section \ref{sec:forecasting} focuses on forecasting the imbalance and the open-loop Area Control Error (ACE). Conclusions and future perspectives are presented in Section \ref{sec:cls}. Additionally, Appendix \ref{appendix-lit} offers a brief literature review on system imbalance analysis and forecasting. 
Finally, appendix \ref{appendix-res} provides details on the data collected and additional results and presents the Figures supporting the explanations of the result sections. 

\section{Problem Statement} \label{pbst}
% méthodologie cf section 3 de la NT + 4.1

This study attempts to determine the explanatory factors for the two variables: the open-loop ACE and the system imbalance.
% Def du déséquilibre sans les activations MA/TERRE et aFRR
This study defines the system imbalance as the open-loop ACE minus the activations on the French balancing mechanism\footnote{The balancing mechanism allows RTE to dispatch tertiary reserves to ensure the generation-consumption balance in real-time, to contribute towards managing network congestion and to reconstitute reserves and frequency ancillary services.} and the activations on the European RR platform TERRE\footnote{\url{https://www.entsoe.eu/network_codes/eb/terre/}} (Trans European Replacement Reserves Exchange). 
% Def de l'open-loop ACE
The open loop ACE for a control area is the ACE\footnote{The ACE is the instantaneous difference between the actual and the reference value for the power interchange of a control area, taking into account the effect of the frequency deviation for that control area.} for that control area minus the automatic Frequency Restauration Reserve (aFRR) activations. It represents the remaining imbalance in the system to deal with automatic reserves once offers have been activated on the balancing mechanism and the TERRE platform.

The variables of interest that may serve as potential explanatory factors are denoted as X. 
Several methods can be used to identify relationships between X and the system imbalance/open-loop ACE: i) calculating correlation coefficients; ii) creating scatter plots of the system imbalance/open-loop ACE against X to observe any potential patterns (e.g., a Gaussian distribution); iii) employing Kernel Density Estimation to assess the density of the system imbalance/open-loop ACE in relation to X; iv) utilizing statistical models (such as ARMA) or machine learning techniques to predict the system imbalance/open-loop ACE based on X.

% Data and explanatory variables
The explanatory variables examined in the following sections include: i) lags of imbalance/open-loop ACE from several periods, ranging from a few minutes to one day ahead; ii) forecast errors for consumption, wind power, and PV generation (forecasted amounts minus actual observations) one hour before delivery and one day in advance; iii) actual consumption, wind power, and PV generation realizations.
Appendix \ref{appendix-data} provides details on the data collected.

\section{System imbalance analysis}\label{sec:imb-analysis}

This section presents the distributions of the imbalance/open-loop ACE vs. explanatory variables. The open-loop ACE autocorrelation analysis is conducted in appendix \ref{appendix:ACF}. 
The Figures of the imbalance and open-loop ACE analysis in Sections \ref{sec:imb-analysis-1} and \ref{sec:imb-analysis-2} are provided in Appendix \ref{appendix-figures}.

\subsection{PV power, wind power and consumption observations explanatory variables}\label{sec:imb-analysis-1}

% Résultat des boxplots déséquilibre/ACEol vs facteur de charge PV/EO/conso
This section aims to identify and, if possible, interpret the evolution of the distributions on average and in terms of quantile deviation by quantifying the width distribution deviation by making the difference between the 99th and 1st quantiles ($\Delta Q = Q_\text{99} - Q_\text{1}$). The results of $\Delta Q$ are provided in appendix \ref{appendix-figures}
Notice that in this section, we refer to the absolute value of $Q_\text{1}$ (since it is negative) when interpreting our results.
Indeed, the French required margin sizing methodology \cite{eem2023dumas} is based on a 1\% security risk criterion. This methodology relies on continuous forecasts of the primary factors contributing to the uncertainties in system imbalance. Four types of forecast errors, which are assumed to be independent, are considered: errors in wind and photovoltaic power generation, errors in the production of conventional power units, and errors in electricity consumption. The required margin is determined by comparing the overall forecast error, calculated as the convolution of the 1\% and 99\% quantiles of these independent errors.
Figure \ref{fig:deseq_ACEol_by_fc_PV_EO_boxplot} depicts boxplots of the imbalance and the open-loop ACE distributions categorized by bins of PV and wind power production load factors (LF) and normalized consumption. 
% Résultats PV
Figure \ref{fig:deseq_ACEol_by_fc_PV_boxplot} shows that both imbalance and the open-loop ACE tend to increase on average with the PV capacity factor, which is expected. 
On average, the imbalance tends to be positive when the load factor exceeds 0.4. This is expected, as such conditions usually indicate overproduction in the system. These situations typically occur during the summer months when PV production is at its highest and consumption is at its lowest, leading to excess production. Activating offers on the French balancing mechanism and on the TERRE platform helps mitigate this imbalance, bringing average imbalance values closer to zero. However, the open-loop ACE remains slightly positive at high load factors, meaning that aFRR is still needed to solve the imbalance. Conversely, for PV load factors below 0.3, the imbalance is slightly negative on average. Low PV production generally occurs in the winter, when system consumption is higher, resulting in a more frequent negative imbalance.
The 99\% percentile of the open-loop ACE and imbalance tends to rise as the PV LF increases, which is a logical outcome of the mean of the distribution shifting toward positive values. Conversely, the 1\% percentile decreases in absolute value as the PV LF increases. 
% Delta_Q PV FC en annexe

% Résultats EO
Figure \ref{fig:deseq_ACEol_by_fc_EO_boxplot} illustrates that the mean values of the imbalance and the open-loop ACE distributions do not show significant variation with changing wind power load factor values. However, the dispersion, represented by the difference between the 75th and 25th percentiles ($Q_\text{75}$ - $Q_\text{25}$), increases for the imbalance distribution. A similar, though less pronounced, trend is also observed for the open-loop ACE.
Unlike PV, wind generation is less seasonally dependent; thus, the system may experience shortages or surpluses in both summer and winter due to low or high load factors. This lower correlation between consumption and wind generation might explain why the average imbalance remains relatively stable regardless of the wind power LF.
Furthermore, both the 1\% and 99\% percentiles of the open-loop ACE and imbalance tend to rise with an increasing wind power LF, indicating greater distribution dispersion. 
% Delta_Q EO FC en annexe

% Résultats conso
Figure \ref{fig:deseq_ACEol_by_fc_conso_boxplot}  illustrates that the distribution of imbalance tends to average negative values when consumption is high (C $>$ Q75 = 0.4, where the consumption is normalized) and positive values when consumption is low (C $<$ Q25 = 0.2). These scenarios indicate that the system becomes more susceptible to imbalance due to either underproduction or overproduction. Conversely, the distributions of the open-loop ACE with consumption levels tend to cluster around zero.
Moreover, the disparity in imbalance increases (the gap between Q75 and Q25 widens by 300 MW) as consumption rises, which is not observed with the open-loop ACE. These differences can be attributed to the activations on the French balancing mechanism and the TERRE platform, which help correct imbalances during high or low consumption periods (upward and downward activations, respectively). This can explain why the average open-loop ACE is close to zero.
The tendency for imbalances to shift towards negative or positive values as consumption increases or decreases is anticipated. The system is better equipped to manage consumption around its median value (between the 25th and 75th percentiles) compared to situations of high consumption (where there are limited means to increase supply) or low consumption (where there are limited means to decrease supply). Consequently, the system is more likely to experience negative imbalances during periods of high consumption and positive imbalances during periods of low consumption.
% Delta_Q conso FC en annexe

% recapitulatif des résultats
In conclusion, regardless of the situations of PV and wind generation and consumption, activations on the French balancing mechanism and the TERRE platform help reduce imbalances. This is evidenced by the fact that the open-loop ACE is, on average, closer to zero than the overall imbalance.
The average distribution of imbalance and open-loop ACE tends to increase with the PV load factor, which is as expected. However, the difference between the 1st and 99th percentiles of imbalance and open-loop ACE shows little variation with changes in the PV load factor.
The wind load factor has minimal impact on the average distribution of imbalance and open-loop ACE. Still, the gaps between the 75th and 25th percentiles and the 99th and 1st percentiles increase significantly.
Generally, the average imbalance distribution is positive at low consumption levels and decreases to negative values as consumption increases, which aligns with expectations. Additionally, the deviation between the 1st and 99th percentiles of imbalance grows with higher consumption. While the average distribution of open-loop ACE remains relatively unaffected by consumption, there is a slight increase in the gap between the 1st and 99th percentiles.

\subsection{PV power, wind power, and consumption forecast errors explanatory variables}\label{sec:imb-analysis-2}

% Def de l'erreur de prédiction
In this section, we define at time $t$ (t corresponding to one hour) the one hour-ahead forecast error $X^\text{err}_t$ as the forecasted value $\hat{X}_{t-1|t}$ (issued at time $t-1$ for $t$) minus the observation $X_t$ for the considered explanatory variable $X$ (consumption, PV, and wind power production): $X^\text{err}_t = \hat{X}_{t-1|t} - X_t$.
Thus, a positive forecast error ($X^\text{err}_t > 0$) indicates that the observed value is lower than the forecasted value. 
For example, if the forecast error for consumption is positive, it suggests that actual consumption is less than expected, potentially leading to excess production in the system.
Similar to the previous section, this section aims to analyze and interpret how the distributions of explanatory variables evolve, both on average and in quantile deviation, calculated as $\Delta Q = Q_\text{99} - Q_\text{1}$. 
The results of $\Delta Q$ are provided in appendix \ref{appendix-figures}.
It is important to note that when interpreting our results, we also consider the absolute value of  $Q_\text{1}$ since it is negative.

% Résultat des boxplots déséquilibre/ACEol par bin des erreurs de prév à échéance 1 h PV/EO/conso
Figure \ref{fig:deseq_ACEol_by_fc_PV_EO_conso_err_boxplot} depicts boxplots of the imbalance and the open-loop ACE by bins of PV, wind power, and consumption one hour-ahead forecast errors. 
%
% Résultats PV
Figure \ref{fig:deseq_ACEol_PV_bin_err_boxplot} illustrates that the imbalance distribution varies on average due to errors in PV forecasts. However, this observation is slightly less pronounced for the open-loop ACE. Overall, no clear correlation is observed. When there are negative PV production errors (indicating more production than expected), the average imbalance is positive, consistent with our observations. We note that the imbalance tends to be negative on average for errors ranging from 1000 to 2500 MW, indicating that the system generally has less production than anticipated. Conversely, when the errors exceed 2500 MW, the imbalance turns positive, contrary to our expectations that it would remain negative.
% Delta Q pv ERR results in appendix

% Résultats EO
Figure \ref{fig:deseq_ACEol_EO_bin_err_boxplot} illustrates the distributions for wind forecast errors. The distribution of the imbalance generally decreases from negative to positive errors. However, for errors greater than 1500 MW, the distribution increases. 
On average, the imbalance tends to be positive for negative forecast errors, indicating that the system is likely in a state of overproduction. The imbalance becomes positive for forecast errors greater than 2000 MW, which does not align with the system being in underproduction, where a negative imbalance would be anticipated. Conversely, the imbalance is generally negative for wind power forecast errors between 500 and 2000 MW, suggesting that the system is experiencing underproduction. The open-loop ACE average value decreases with an increasing wind forecast error, which is closer to expectations.
%
% Quantiles results
The open-loop ACE's 99th percentile typically decreases from negative to positive errors, while the 1st percentile tends to increase. 
% Delta Q EO erreur prévision results in appendix

% Résultats conso
Figure \ref{fig:deseq_ACEol_conso_bin_err_boxplot} illustrates the boxplots for consumption errors.  
The open-loop ACE tends to increase slightly with positive errors. The average imbalance distribution also tends to rise with the error, mainly when the errors are positive. This pattern is consistent with the observation that a positive consumption forecast error indicates that the actual consumption is less than the forecasted amount. In this situation, the system experiences a production surplus, resulting in a positive imbalance. 
The 99th percentile of the open-loop ACE progressively increases and becomes significant for positive errors greater than 1000 MW. Conversely, the 1st percentile decreases in absolute value as it moves from negative to positive errors. However, relatively few open-loop ACE values exist for errors exceeding 1000 MW, making it challenging to draw any confident conclusions in this error range. 
% Delta Q EO erreur conso results in appendix

% ACEol/déséquilibre vs forecast errors interpretations 
The differences between the open-loop ACE and the general imbalance trend can be understood by examining how the imbalance is corrected through activations on the French balancing mechanism and the TERRE platform. When the forecasts for wind, PV generation, or consumption (positive or negative) significantly deviate from actual outcomes, RTE takes measures to absorb the imbalance. 
Figure \ref{fig:deseq_ACEol_by_fc_PV_EO_conso_err_boxplot} illustrates that, on average, the open-loop ACE includes these corrections, resulting in values closer to 0 compared to the imbalance. This is especially evident when there are negative errors, such as when wind or PV generation is lower than expected or when consumption is higher than anticipated. The effect is somewhat less pronounced for positive errors, where the system experiences an overproduction.
This can be explained by the fact that it is easier to balance the system when there is a shortfall in generation (a lack of production) through upward activations rather than when there is excess generation (overproduction), which requires downward activations. The grid and generation facilities have historically been designed to accommodate peak consumption. However, in recent years, the rapid growth of PV and wind power capacity has caused a shift in this paradigm, leading the grid to encounter overproduction situations increasingly.
Despite this shift, PV and wind power plants still play a minor role in the French balancing mechanism and the TERRE platform, even though they hold the potential to help reduce activation needs.

% conclusions
In conclusion, the boxplots and the 1\% and 99\% percentiles do not clearly indicate any notable correlations in explanatory variables to interpret the imbalance and open-loop ACE trends. 
The differences between the open-loop ACE and the imbalance concerning general trends can be explained. The distributions of the open-loop ACE tend to be closer to zero than the imbalance, indicating that the activations in the French balancing mechanism and TERRE have been effective. 
Moreover, the imbalance tends, on average, to be positive when consumption forecast errors are significant in the positive values (less consumption realized than forecasted, so the system is in excess of production since the error is defined as forecast minus realization), which is consistent. 
The imbalance tends to be positive when the forecast errors for wind and PV production are negative (meaning more production is realized than forecasted). Conversely, when the forecast errors are positive, the imbalance is negative, which aligns with expectations, indicating that the system tends to be short on production.
The only exception to this trend occurs when PV and wind generation forecast errors exceed 2500 MW, where the imbalance is positive despite expectations of a negative imbalance. This may be partly due to the effects of negative spot prices. In such circumstances, RES producers are encouraged to stop their production, but they don't always. Hence, large forecast errors in PV and wind generation arise because RTE struggles to predict which RES power plants will shut down, as wind and PV producers do not always disclose their production plans to RTE. Consequently, the system becomes more complex to balance, leading to anticipation of more shutdowns than actually happen, which results in an overall surplus of production with positive forecast errors.

\section{Machine learning model}\label{sec:forecasting}
% Section 4.5

This section studies how a machine learning model can accurately predict system imbalance/open-loop ACE and identify its explanatory variables.
The machine learning model used in this study is the Gradient-Boosted Decision Trees type, specifically the \textit{HistGradientBoostingRegressor} (HGBR) model from the scikit-learn Python library \cite{scikit-learn}.
The studied dataset spans approximately one year, from May 2022 to July 2023. Appendix \ref{appendix:ML} presents the training and evaluation methodology, the results of the metric values, and the Figures. 

There are two categories of variables considered: 
% description des variables considerées
1) Variables available one day ahead, which include day-ahead forecasts of wind power, PV generation, and consumption, as well as the 24-hour lags of the imbalance/open-loop ACE;
2) Variables available close to real-time, such as wind power, PV generation, and consumption forecasts, provided 1 or 2 hours before delivery, along with 1-hour and 2-hour lags of the imbalance/open-loop ACE.
Therefore, the inputs (explanatory variables) used in the HGBR model include:
X1: Realized wind power, PV generation, and consumption;
X2: Imbalance/open-loop ACE lags from 5 to 60 minutes in 5-minute intervals (lags close to real-time);
X3: Imbalance/open-loop ACE lags from 23 to 25 hours in 5-minute increments, along with day-ahead wind power forecasts, PV generation, and consumption.
It is important to note that X3 includes the day-ahead variables that could be utilized as inputs for the operational mFRR sizing model.
Indeed, in practice, in the French system, mFRR is contracted in D-1, so the mFRR's sizing can only incorporate the variables available on a day-ahead basis.

% training explanations
A single HGBR model is trained to forecast the mean value of the imbalance/open-loop ACE, and an additional model is created for each percentile—1\%, 50\%, and 99\%—with identical hyperparameters across all models. This results in a total of four models. 
% metrics considered explanations
The quality of the mean forecast is evaluated using the Mean Absolute Error (MAE) and Root Mean Square Error (RMSE) metrics. In contrast, the Pinball Loss (PL) metric is employed for the 1\%, 50\%, and 99\% percentiles. Lower scores indicate better model performance for each metric. 

\subsection{System imbalance forecasting}\label{sec:imb-forecasting}

Table \ref{tab:imb-ace-results} presents the average values per metric over the four splits from the 4-fold cross-validation. Each split yields a value for each score.
%
% Interpretations des résultats du tableau
The X2 variable, which consists of imbalance lags from 5 to 60 minutes before delivery, captures nearly all the predictive power of imbalance. This is especially true for the 5-minute and 10-minute lags. However, the PL values for the 1st and 99th percentiles are relatively good when only considering the day-ahead variables (X3), compared to results obtained with X1 or X2 inputs. This finding may be attributed to the fact that predicting values at the tails of the distribution is generally easier as these percentiles are more conservative.
% Interpretations des résultats de la Figure 
Figure \ref{fig:deseq_prediction} illustrates the predictions of the 1st percentile (red), the 50th percentile (green), the 99th percentile (purple), and the mean value (black) of the imbalance, in comparison to the actual observations (blue) from a specific day in the testing set. It compares the predictions of the HGBR model using the X1+X2+X3 inputs to the models with X2 and X3 inputs only.
It is evident that there is a significant difference between the model that includes X2 as an explanatory variable (shown in the top and middle sections of the figure) and the model that relies exclusively on X3 (shown at the bottom). The model utilizing X3 as inputs, which represents day-ahead variables, can only predict a relatively constant "envelope" of the imbalance In contrast, the models incorporating recent lags of the imbalance (X1 and X2) achieve much more accurate predictions. This result highlights the challenges of predicting imbalance with high accuracy on a day-ahead basis.
% Interpretations des résultats de la Figure avec scatter plot
Figure \ref{fig:deseq_prediction_scatter} shows the relationship between realized and predicted imbalances in the testing set. An "accurate" model would produce predictions that closely align with the actual realizations, resulting in a scatterplot that nearly overlaps a straight line (shown in red on the graph). However, when lags of 5 to 60 minutes are excluded from the explanatory variables, the model struggles to predict imbalances accurately on average.

% conclusions
In conclusion, predicting the mean or median value of the imbalance distribution for the following day is challenging. However, forecasting the 1st and 99th percentiles is more manageable, as noted in \cite{DEVOS2019272}.
Recent lags, ranging from 5 to 120 minutes of real-time data, can help predict the imbalance. Unfortunately, this explanatory variable is not currently considered in the sizing of the mFRR, which is determined on a day-ahead basis. Nonetheless, we could envision incorporating these recent lags into RTE's imbalance forecast for short lead times (i.e., 2 hours) to better anticipate imbalances and optimize the activations on the balancing mechanism and the TERRE platform.
Additionally, this analysis could be enhanced by including producer production plans for thermal generation as an explanatory variable in the imbalance forecasting model.

\subsection{Open-loop ACE forecasting}\label{sec:ace-forecasting}

\begin{figure}[tb]
    \begin{subfigure}{.5\textwidth}
    \centering
    \includegraphics[width=\linewidth]{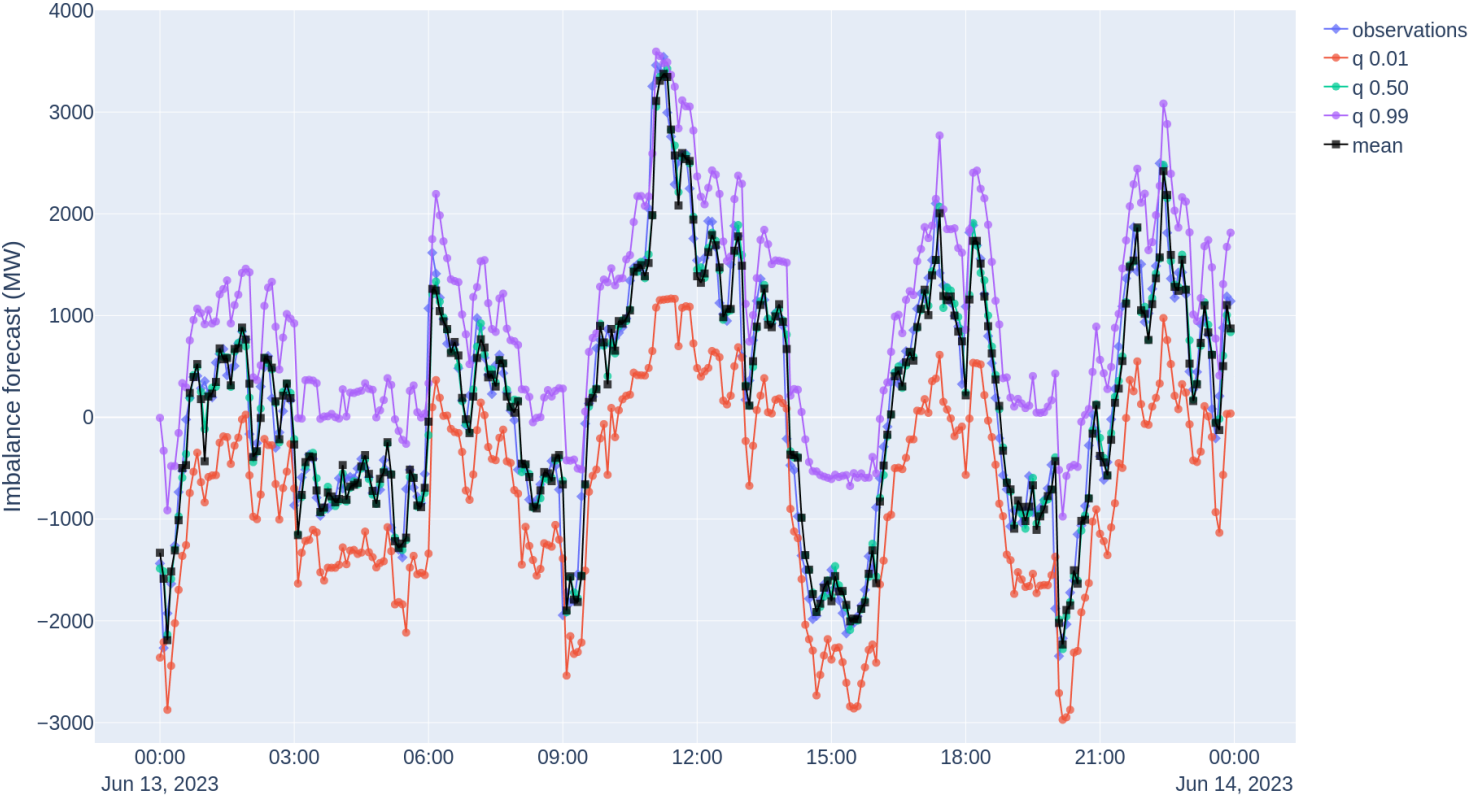}
    \caption{X1+X2+X3 inputs.}\label{fig:deseq_X1X2X3_prediction}
    \end{subfigure}
    \begin{subfigure}{.5\textwidth}
    \centering
    \includegraphics[width=\linewidth]{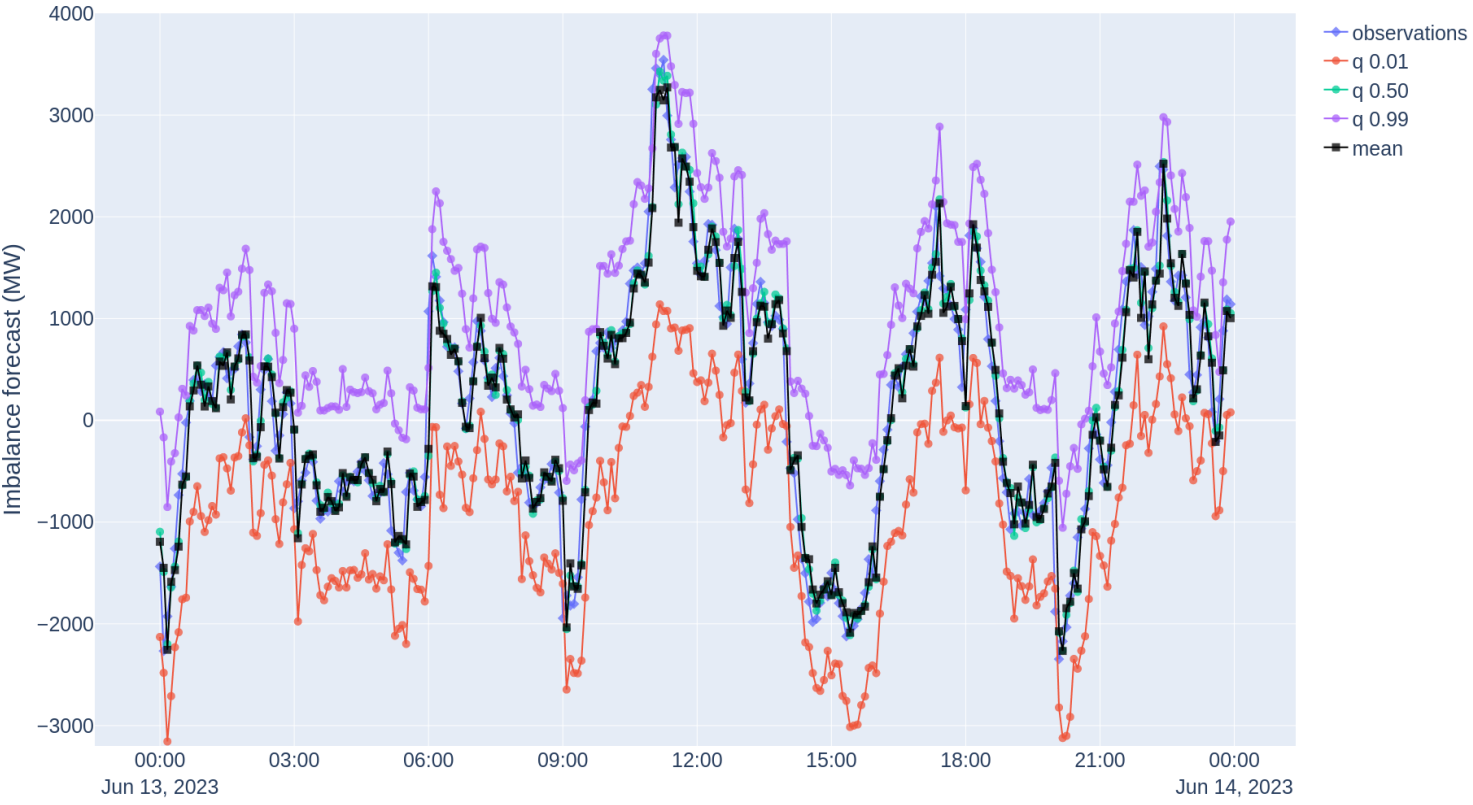}
    \caption{X2 inputs.}\label{fig:deseq_X2_prediction}
    \end{subfigure}
    \centering
    \begin{subfigure}{.5\textwidth}
    \centering
    \includegraphics[width=\linewidth]{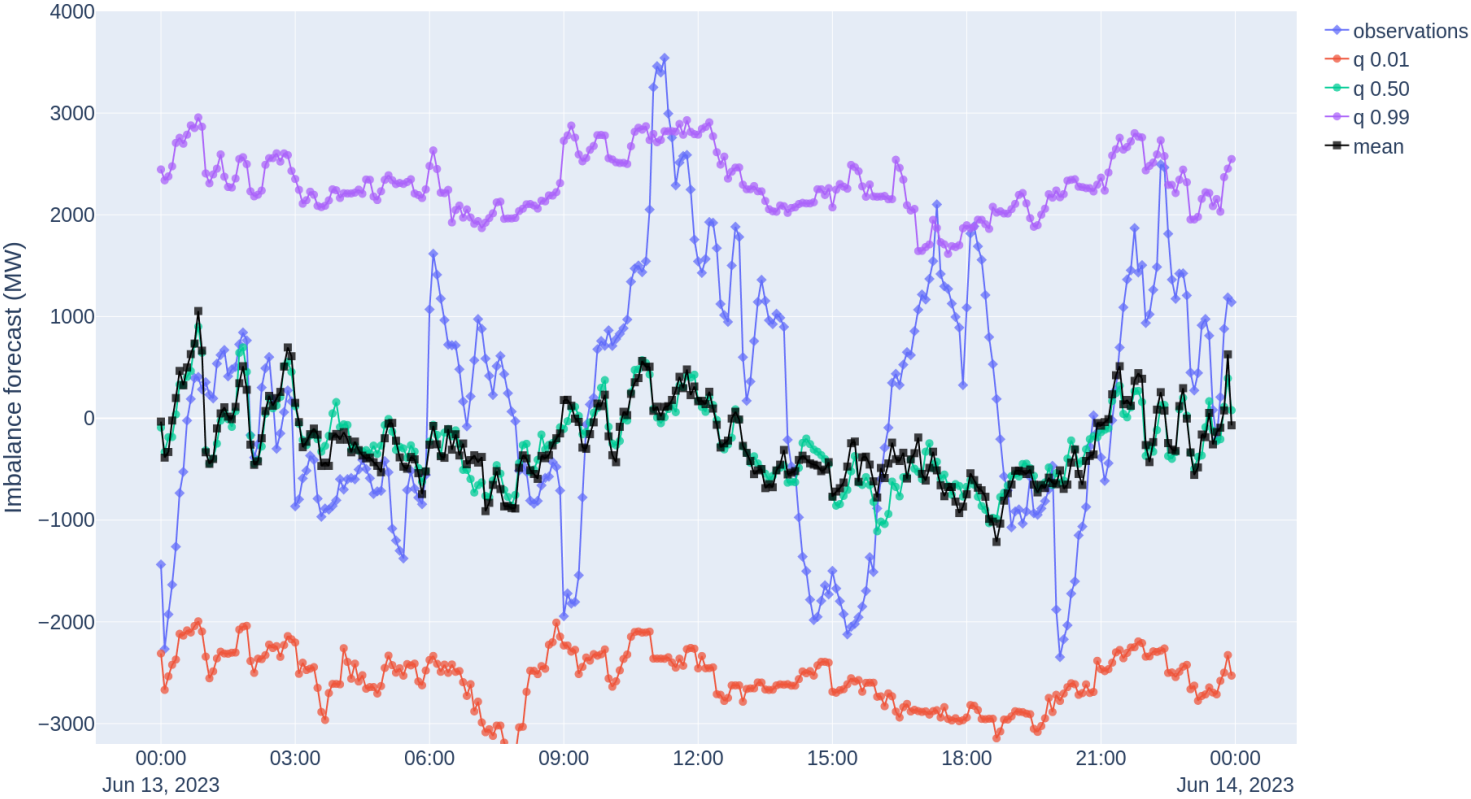}
    \caption{X3 inputs.}\label{fig:deseq_X3_prediction}
    \end{subfigure}
    \caption{The predictions of the 1st percentile (red), the 50th percentile (green), the 99th percentile (purple), and the mean value (black) of the imbalance, in comparison to the actual observations (blue) from a specific day in the testing set.}\label{fig:deseq_prediction}
\end{figure}

\begin{figure}[tb]
    \begin{subfigure}{.5\textwidth}
    \centering
    \includegraphics[width=\linewidth]{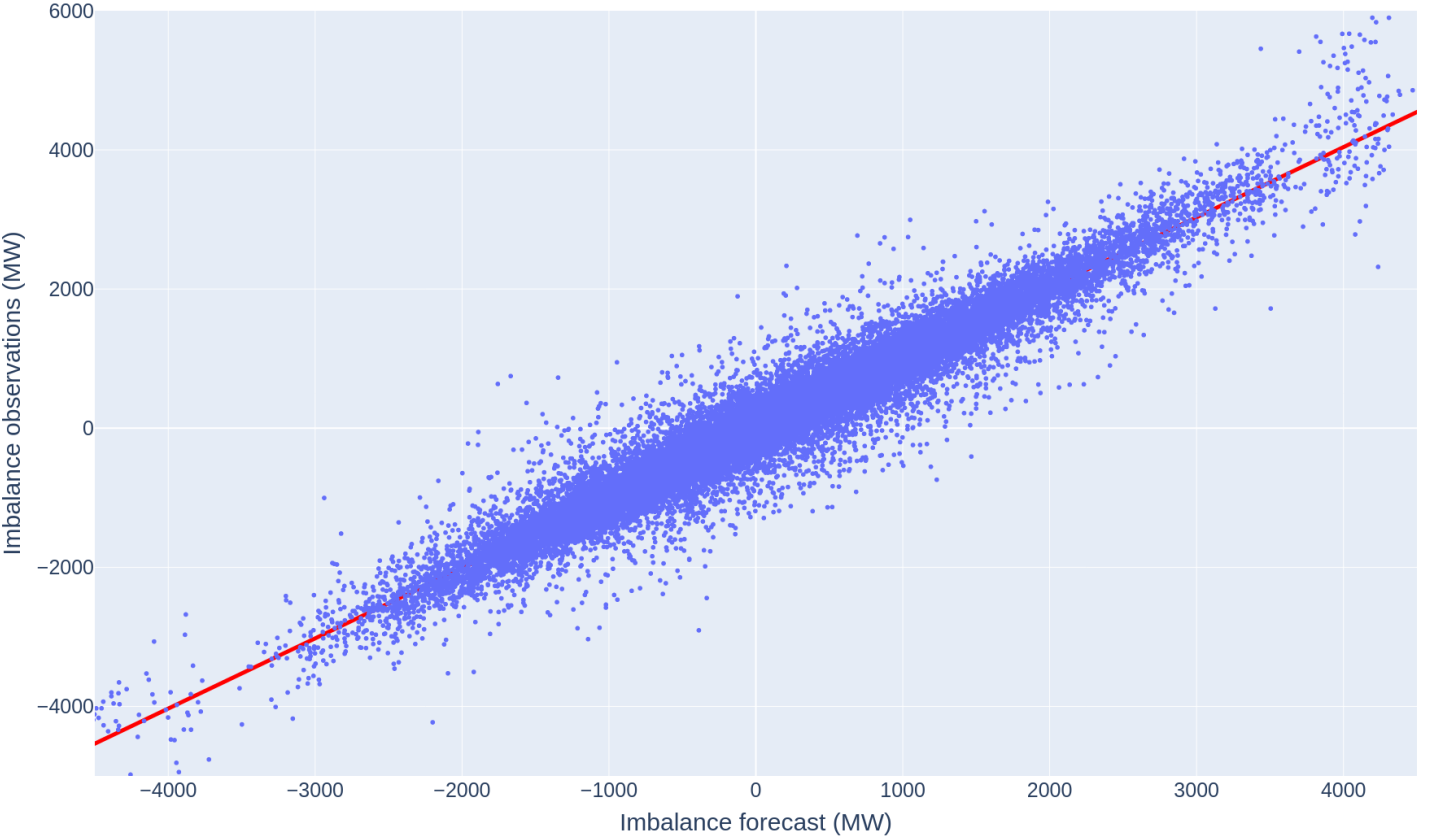}
    \caption{X1+X2+X3 inputs.}\label{fig:scater_X1X2X3_deseq}
    \end{subfigure}
    \begin{subfigure}{.5\textwidth}
    \centering
    \includegraphics[width=\linewidth]{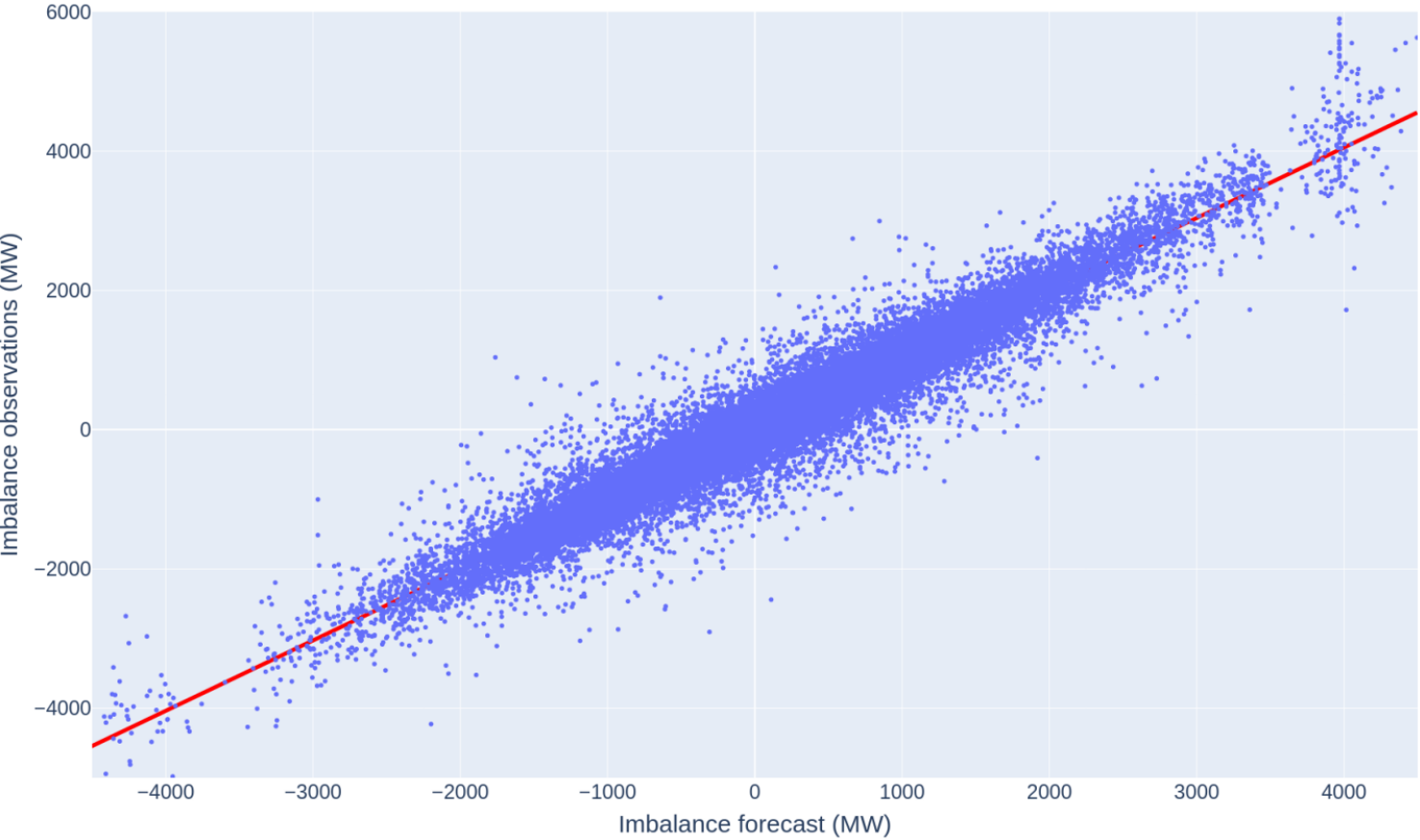}
    \caption{X2 inputs.}\label{fig:scater_X2_deseq}
    \end{subfigure}
    \centering
    \begin{subfigure}{.5\textwidth}
    \centering
    \includegraphics[width=\linewidth]{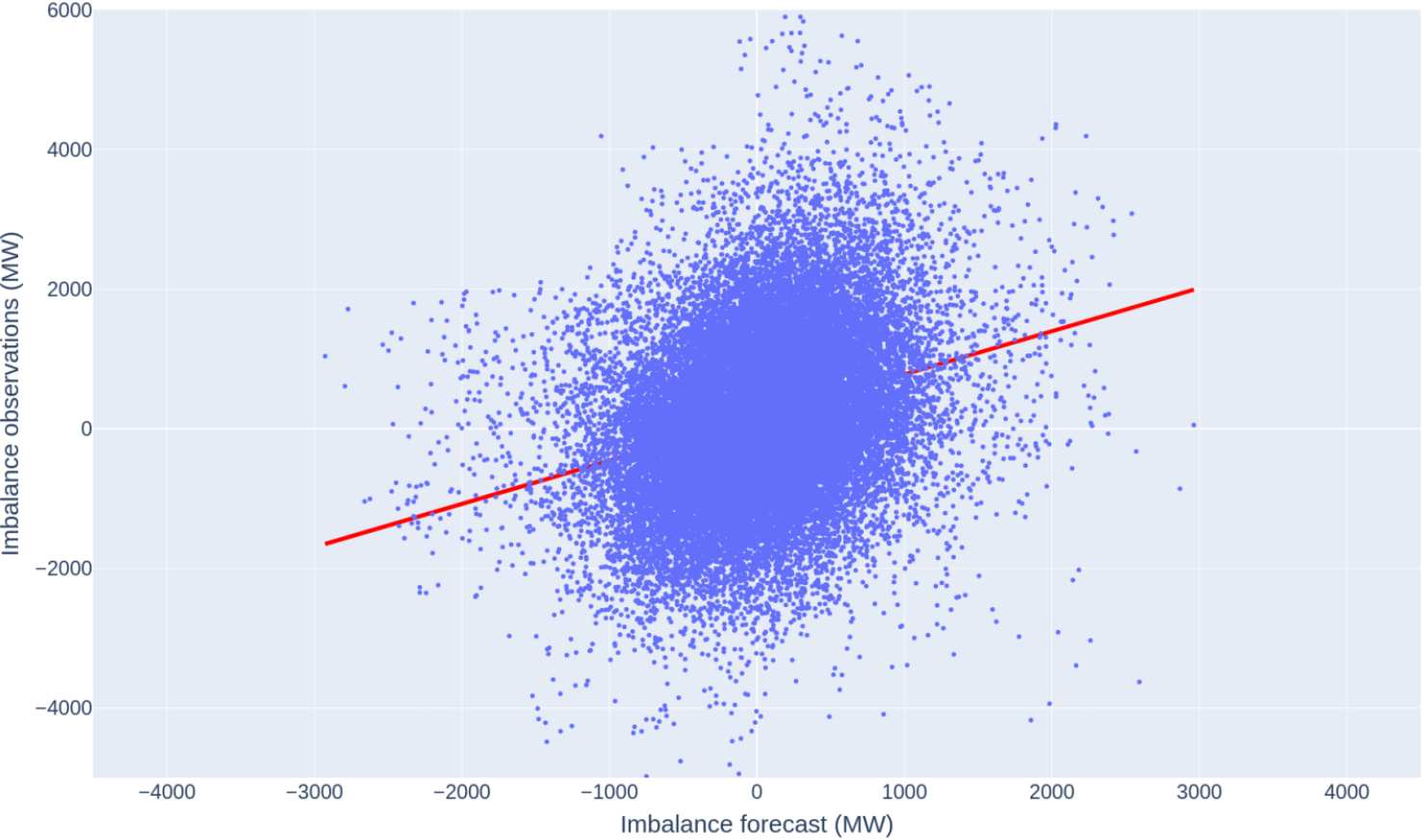}
    \caption{X3 inputs.}\label{fig:scater_X3_deseq}
    \end{subfigure}
    \caption{Prediction of the average imbalance compared to the observations over the testing set.}\label{fig:deseq_prediction_scatter}
\end{figure}

Similar to the previous section \ref{sec:imb-forecasting}, this section aims to predict the open-loop ACE and identify its explanatory variables. 
The lower part of table \ref{tab:imb-ace-results} presents the MAE, RMSE, and PL metrics scores for the open-loop ACE production.
% conclusions 
The conclusions about the explanatory variables are the same as those in the previous section \ref{sec:imb-forecasting} regarding the prediction of imbalance. Thus, we do not include Figures such as \ref{fig:deseq_prediction} and \ref{fig:deseq_prediction_scatter}.
Predicting the mean or median value of the open-loop ACE for the next day is challenging. However, forecasting the 1st and 99th percentiles is more manageable,
It is important to note that the metric values are lower for open-loop ACE forecasting compared to imbalance forecasting. This occurs because the distribution of open-loop ACE is smaller than that of imbalance, as shown in the graphs of section \ref{sec:imb-analysis} as RTE implemented activations on both the balancing mechanism and the TERRE platform to maintain system balance.

\section{Conclusions and perspectives}\label{sec:cls}

% recapitulatif des resultats
The findings presented in section \ref{sec:imb-analysis} indicate that a priori, the imbalance and open-loop ACE show no correlation with wind power, PV generation, or consumption forecasts, as well as the actual realizations of these variables. These results are consistent with \cite{GOODARZI2019110827,kratochvil2016system}. One possible interpretation is that the system imbalance arises from the balancing responsible parties (BRP) imbalance, which cannot be attributed to their forecasts. The BRPs are incentivized to achieve balance up to the last intraday gate, which occurs one to two hours before real-time. Consequently, all upstream variables may be irrelevant. This would also explain why the imbalance and open-loop ACE strongly correlate with their realizations for up to one hour (as noted in appendix \ref{appendix:ACF}) but do not have a significant correlation beyond that timeframe. These findings are further substantiated by the attempts to predict the imbalance and the open-loop ACE in section \ref{sec:forecasting}. 
Moreover, \cite{EICKE2021105455} shows that BRPs also have a strategic behavior and can voluntarily choose to be imbalanced. This possibility has not been considered in this study. It may be worthwhile to consider whether variables such as the intraday price or reserve activations could help explain the imbalances experienced by the BRPs.

The mFRR sizing is performed on a day-ahead basis. Given this timeframe and the results from this study, it currently seems impossible to accurately predict the average imbalance/open-loop ACE using the explanatory factors examined here. However, it is possible that other variables, such as spot and intraday prices, could alter this conclusion. Accurate predictions may be achievable one to two hours before real-time at the beginning of the operational window of the TSO (currently two hours ahead in France). Longer lead times could be achievable if it is feasible to simulate the actions of the BRPs from the day before until the start of the operational window.

The reserve sizing method outlined in \cite{DEVOS2019272} is effective as it accurately predicts the extreme quantiles of the imbalance distribution (specifically the 0.1\% and 99.9\%). Results presented in section \ref{sec:forecasting} indicate that predicting the distribution's tails is easier than predicting the mean or median. This is likely due to the larger interval, which allows for a more conservative estimation. 
Based on this method, performing D-1 reserve sizing using historical data and forecasts would be feasible, similar to what is described in \cite{DEVOS2019272}. The findings in section \ref{sec:forecasting} could potentially be enhanced by identifying similar historical days using a k-nearest neighbors (k-NN) approach. For D-1, we can ascertain the "average" conditions for day D, such as high PV and wind power generation alongside low consumption. These specific conditions may have an impact on the imbalance distribution. However, further investigation is necessary to explore this possibility fully.

% A rajouter après les reviews
%\section*{Acknowledgment}

%TODO remercier Arman pour sa relecture ! 

% references section
\bibliographystyle{ieeetr}
\bibliography{biblio}

\section{Literature review}\label{appendix-lit}

Examining the relationship between system imbalances and predictions of adequacy variables, such as renewable energy production and consumption, can lead to the development of advanced methods for forecasting imbalances. This forecasting could be performed from one day in advance to just a few hours in real-time. Such methods would enable us to accurately determine the necessary amount of operating reserves needed to ensure a secure supply with a high level of confidence. However, as noted in the introduction, there are only a few articles that focus on this topic.
Four articles have served as significant sources of inspiration for the work presented in this paper. This section summarizes these articles and compares their approaches to the methodology used in this article.

% Dimensionnement dynamique en Belgique
The study \cite{DEVOS2019272} presents a method for dynamically sizing operating reserves using a machine learning (ML) model to predict the quantiles of the probability of system imbalance, specifically the open-loop ACE. The explanatory variables include renewable energy production (from wind and PV sources), energy consumption, and outages of thermal power plants. The model is trained on historical data and is executed daily, using day-ahead forecasts of these variables, along with calendar information and broader data such as day-ahead weather forecasts.
The ML model processes this information to predict the 0.1\% and 99.9\% quantiles of the probability density of the open-loop ACE. These quantiles determine the necessary operating reserves that need to be contracted. 

Two machine learning models are employed: a K-means model and a k-nearest neighbors (k-NN) model. They are evaluated using a case study based on the Belgian power system, demonstrating greater efficiency than the previously used statistical approach.
This study demonstrates the potential to predict extreme quantiles of the probability distribution of future imbalances one day in advance, using available forecast data. These findings align with the conclusions drawn in the paper: while accurately predicting future imbalances is difficult, obtaining a reasonable approximation of the extreme quantiles of their probability distribution based on forecasts is achievable. This approach could be utilized to develop innovative methodologies for sizing operating reserves.

% Relation dimensionnement/erreur de prévision en Allemagne
The study \cite{GOODARZI2019110827} serves as a valuable complement to \cite{DEVOS2019272} by examining the relationship between imbalances in Germany's energy market and forecast errors related to RES. Additionally, it investigates how these forecast errors impact intraday prices.
The research employs forecast errors, lagged values of the open-loop ACE, and calendar variables as inputs for a multivariate linear regression model. Two regressions are conducted: the first utilizes the Ordinary Least Squares metric as the loss function, while the second applies the pinball loss function. The study concludes that forecast errors significantly affect imbalances. Notably, wind forecast errors have a greater impact on these imbalances than solar forecast errors, primarily because a substantial portion of the imbalance occurs during evening peak hours when solar production is typically low. 

Furthermore, there is a correlation between market prices and excess electricity generation, underscoring another connection between imbalance and market participant behavior. These findings suggest that exploring the relationship between real-time imbalances and market dynamics could be an important area for future research.

% Etude des facteurs explicatifs du déséquilibre en République Tchèque
Similarly, the study by \cite{kratochvil2016system} aims to predict imbalances in the Czech Republic in near real-time. It examines the correlation between these imbalances and various explanatory variables, such as RES production, net position forecasts, forecast errors, and prices in energy markets (including spot, intraday, and balancing markets). However, the correlations identified in this research were insignificant, which aligns with the conclusions reached in this article.

% Création d'un marché fictif du déséquilibre en Allemagne
The necessity of considering actor behavior in future studies is underscored by the research presented in \cite{EICKE2021105455}, which establishes a fictional imbalance market to examine the relationship between imbalance and the strategies of market participants, using data from the German electricity system. The tests conducted on this imbalance market reveal a correlation between the quantity of imbalance and the price of imbalance. This finding highlights the presence of strategic behaviors among actors who may intentionally create imbalances within their scope to maximize their profits, particularly concerning wholesale market prices. This issue is especially noteworthy in Germany, where strategic deviations are theoretically prohibited.

\section{Additional results and explanations}\label{appendix-res}

This appendix details the data collected and additional results and presents the Figures supporting the explanations of the result sections \ref{sec:imb-analysis} and \ref{sec:forecasting}. 

\subsection{Data origin}\label{appendix-data}

The RTE open-loop ACE data is collected with a granularity of 5 seconds and is then resampled to 1-minute and 5-minute intervals for analysis in the following sections. The data from the French balancing mechanism and TERRE activations is available with a granularity of 5 minutes. Observations and forecasts for PV, wind power, and French consumption are provided at a granularity of 30 minutes. The system imbalance is calculated by differentiating between the open-loop ACE resampled at 5-minute intervals and the activations from the balancing mechanism and TERRE. To compare the system imbalance/open-loop ACE with the explanatory variable, we take the value corresponding to the first 5 minutes of each half-hour interval.

\subsection{Autocorrelation of the open-loop ACE}\label{appendix:ACF}
% Section 4.2 -> focus 4.2.1 autocorrelation

\begin{figure}[tb]
    \centering
    \includegraphics[width=\linewidth]{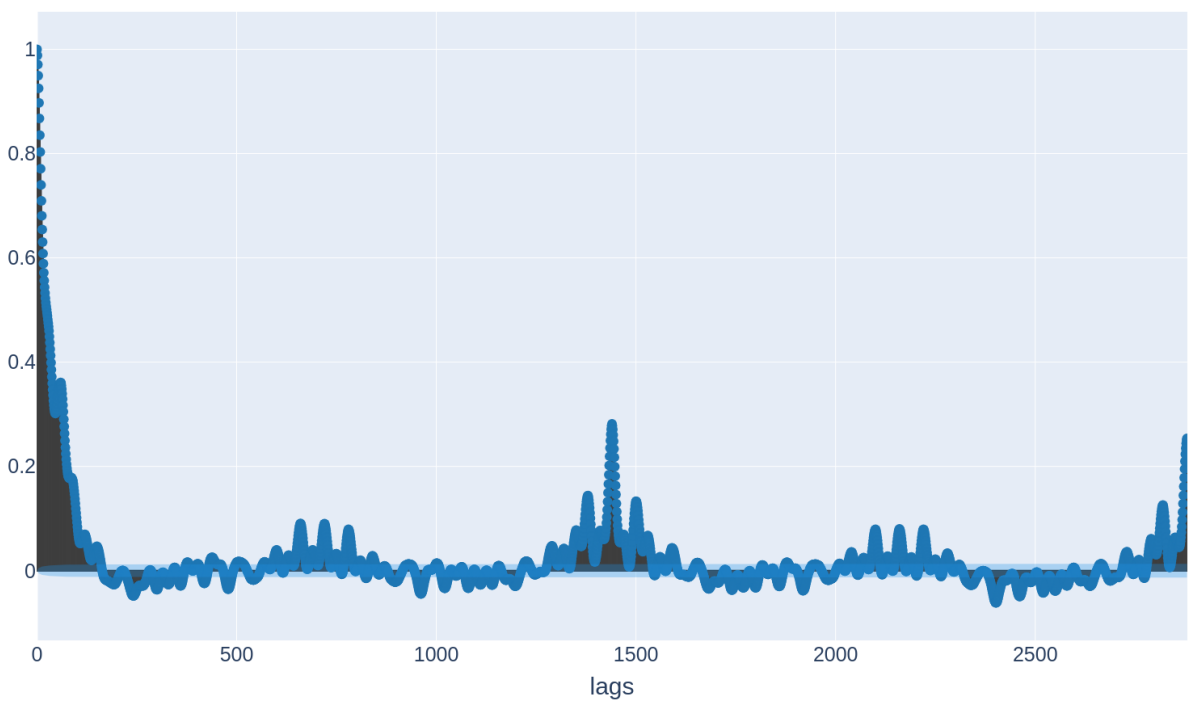}
    \caption{ACF of the open-loop ACE at 1-minute granularity.}
    \label{fig:ACEol_ACF}
\end{figure}

The open-loop ACE autocorrelation analysis is conducted at a 1-minute interval because performing it at a 5-second interval is too time-consuming, and the results are nearly identical. The analysis spans 1440 minutes over two days, resulting in 2880 lags. The results of this autocorrelation function (ACF) are presented in Figure \ref{fig:ACEol_ACF}.
%
%%%%
% ACF du déséquilibre
%%%
%
The findings reveal significant correlations at the following time indexes: 
\begin{enumerate}
    \item Minutes 1 to 90: values greater than 0.3 up to minute 60 and between 0.2 and 0.4 up to minute 90.
    \item Hours 11, 12, and 13: peaks observed at lags of 660 (60*11), 720 (60*12), and 780 (60*13), with values around 0.1.
    \item Hours 23 and 25: peaks at 1380 (60*23) and 1500 (60*25) lags, with values close to 0.15.
    \item Hour 24: a peak at lag 1440 (60*24) with values close to 0.3.
\end{enumerate}
The most significant autocorrelations are achieved with the closest lags ($<$ 1 hour).
%
% Conclusion
These autocorrelation calculations reveal that specific "lags" exhibit significant correlations with the open-loop ACE observations, particularly during the following periods: the first two hours, hours 11 to 13, and hours 23 to 25. It will, therefore, be relevant to include them in Section \ref{sec:imb-forecasting} as input variables for a predictive model.

\subsection{Additional results on the imbalance/open-loop ACE distribution analysis}\label{appendix-figures}

\begin{figure}[tb]
    \begin{subfigure}{.5\textwidth}
    \centering
    \includegraphics[width=\linewidth]{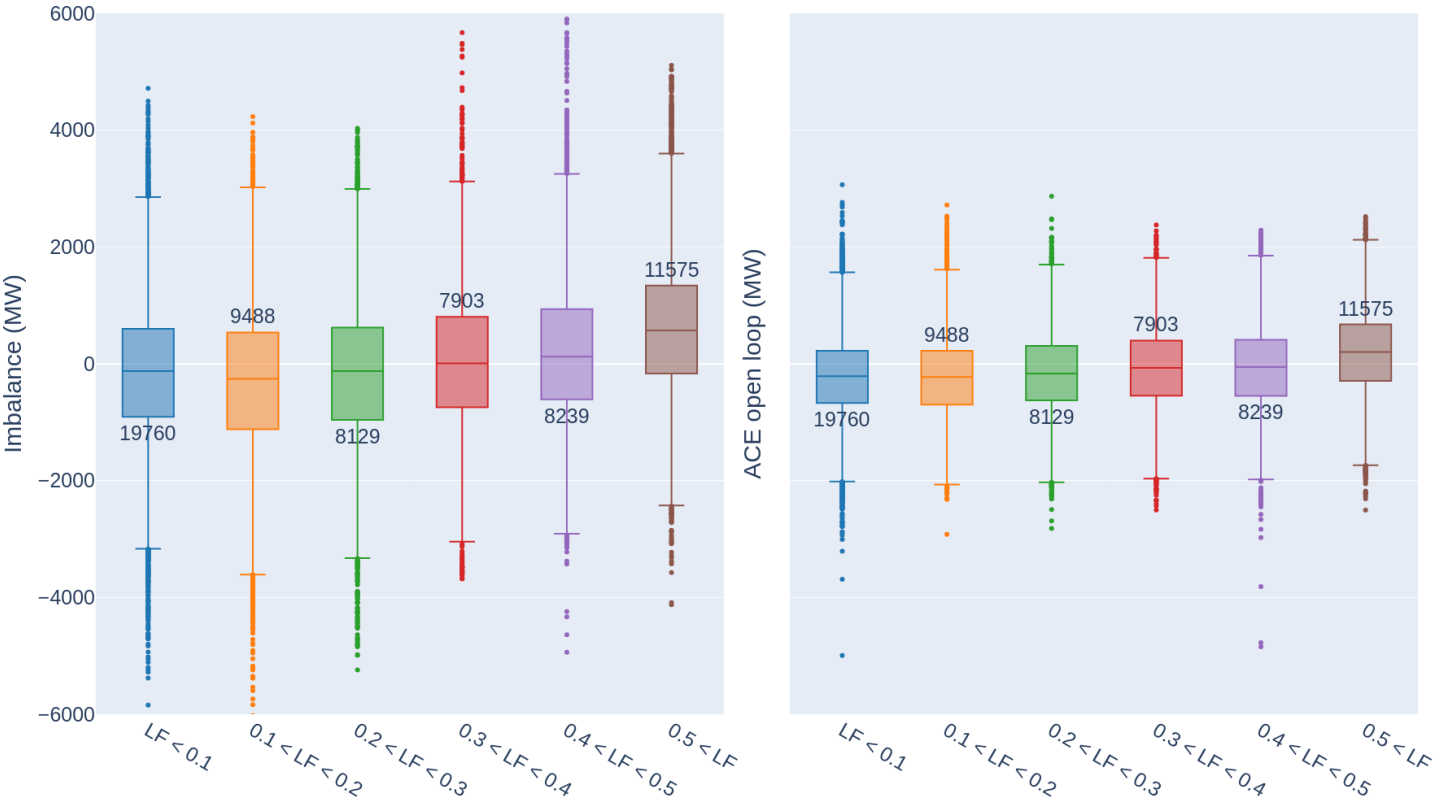}
    \caption{Imbalance (left) and the open-loop ACE (right) by bins of PV LF.}\label{fig:deseq_ACEol_by_fc_PV_boxplot}
    \end{subfigure}
    \begin{subfigure}{.5\textwidth}
    \centering
    \includegraphics[width=\linewidth]{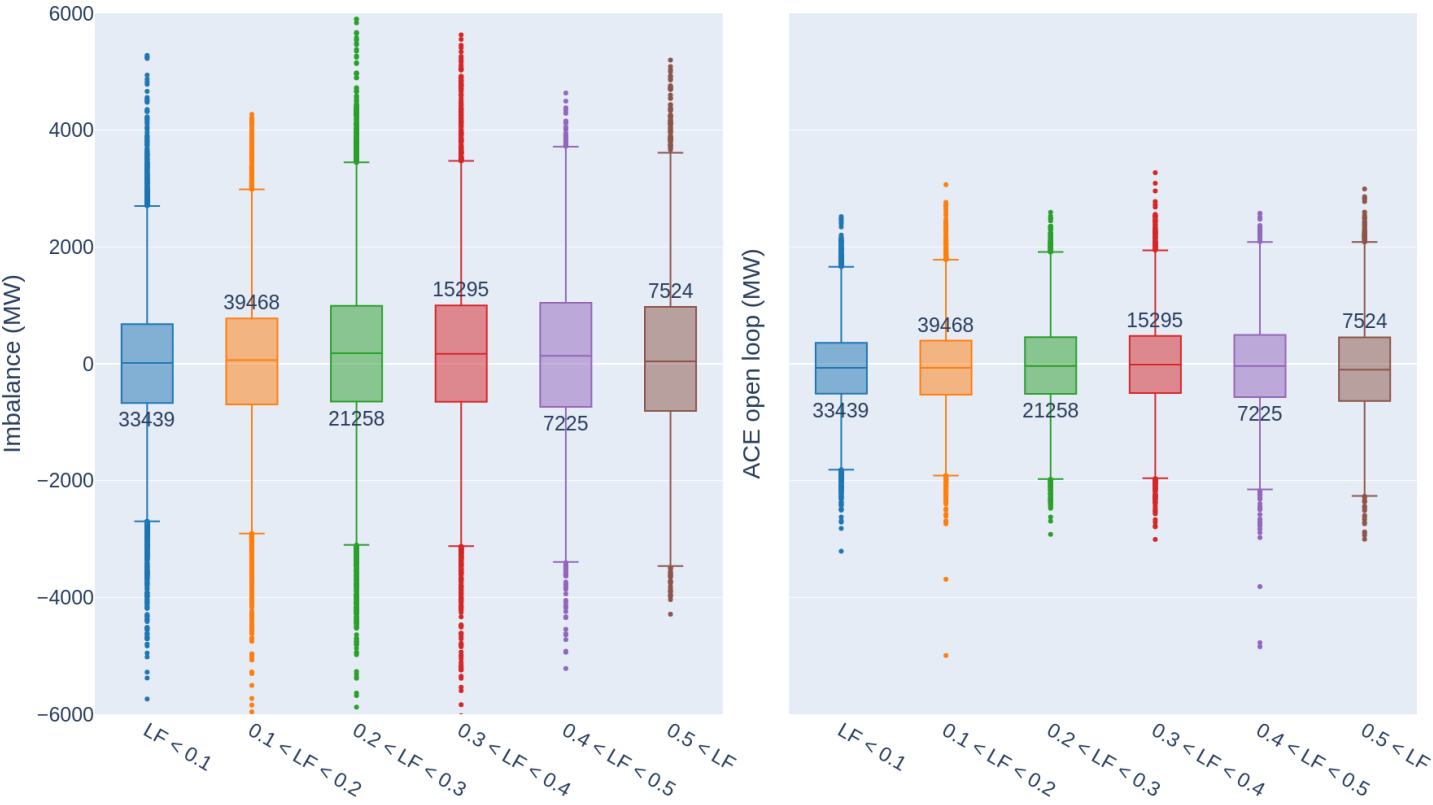}
    \caption{Imbalance (left) and the open-loop ACE (right) by bins of wind power LF.}\label{fig:deseq_ACEol_by_fc_EO_boxplot}
    \end{subfigure}
    \begin{subfigure}{.5\textwidth}
    \centering
    \includegraphics[width=\linewidth]{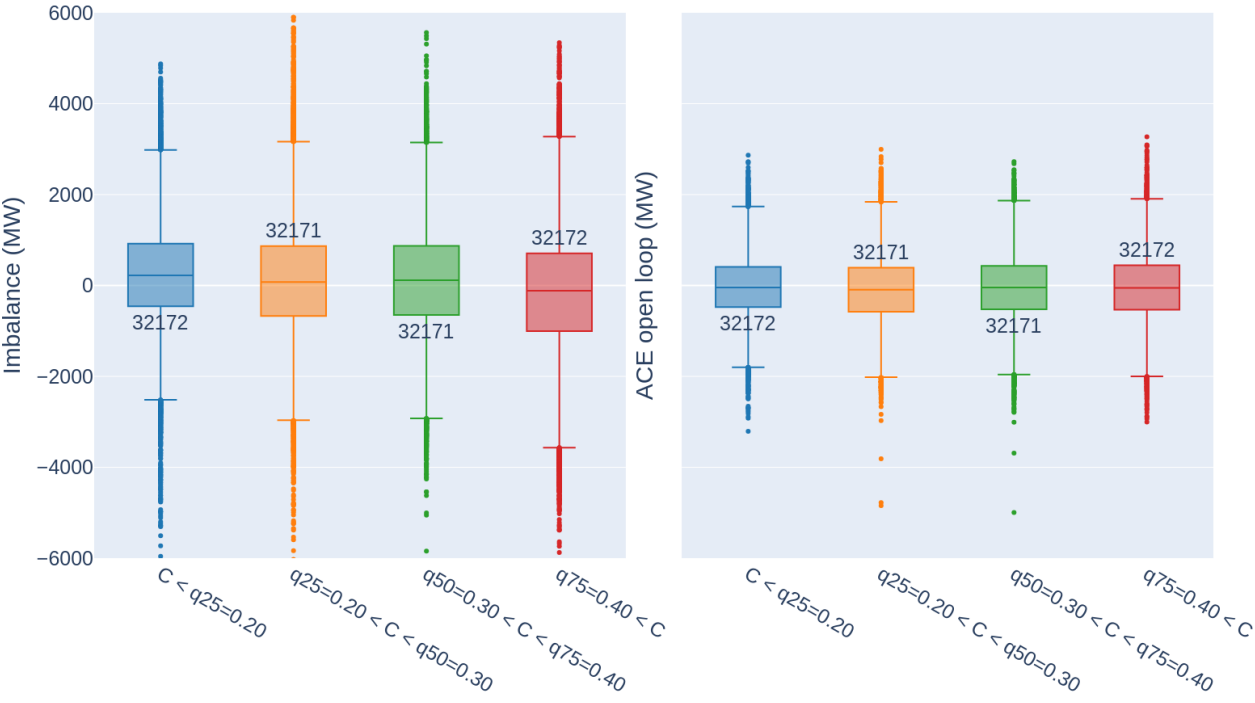}
    \caption{Imbalance (left) and the open-loop ACE (right) by bins of normalized consumption (C).}\label{fig:deseq_ACEol_by_fc_conso_boxplot}
    \end{subfigure}
    \caption{Boxplots of the imbalance (left) and the open-loop ACE (right) by bins of PV (top) and wind power (middle) production load factors (LF), and normalized consumption (bottom). The number of values by boxplot considered is displayed. }\label{fig:deseq_ACEol_by_fc_PV_EO_boxplot}
\end{figure}

\begin{figure}[tb]
    \begin{subfigure}{.5\textwidth}
    \centering
    \includegraphics[width=\linewidth]{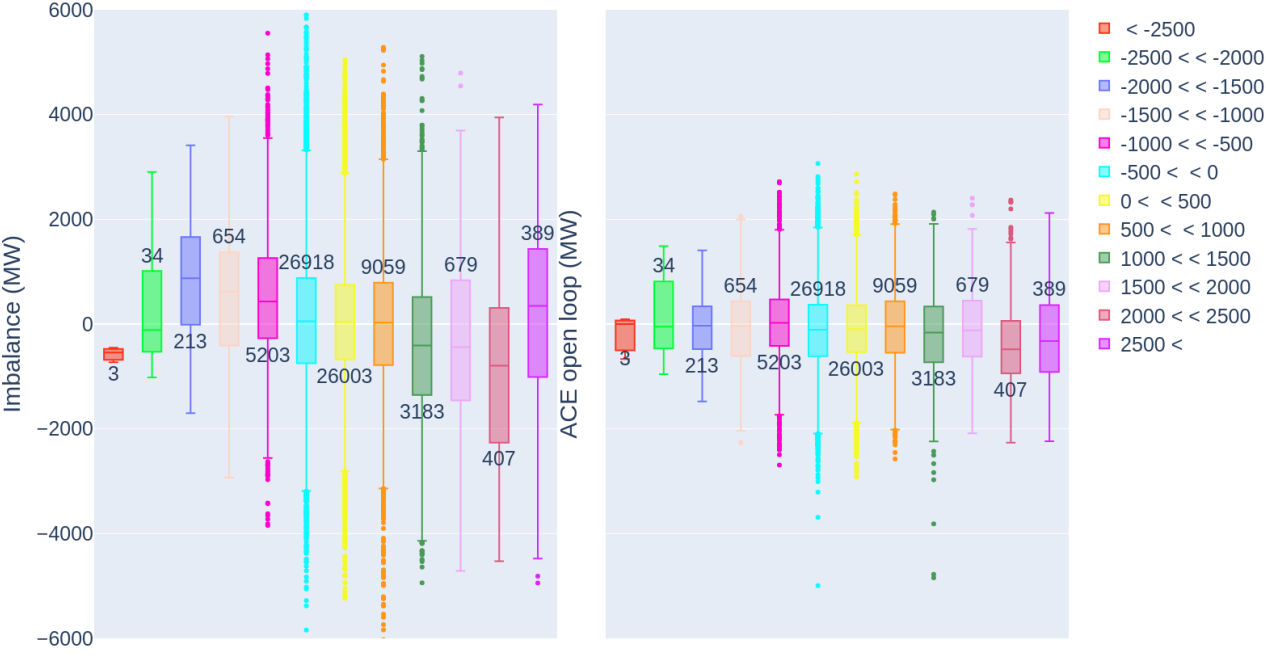}
    \caption{Imbalance (left) and the open-loop ACE (right) by bins of PV forecast errors.}\label{fig:deseq_ACEol_PV_bin_err_boxplot}
    \end{subfigure}
    \begin{subfigure}{.5\textwidth}
    \centering
    \includegraphics[width=\linewidth]{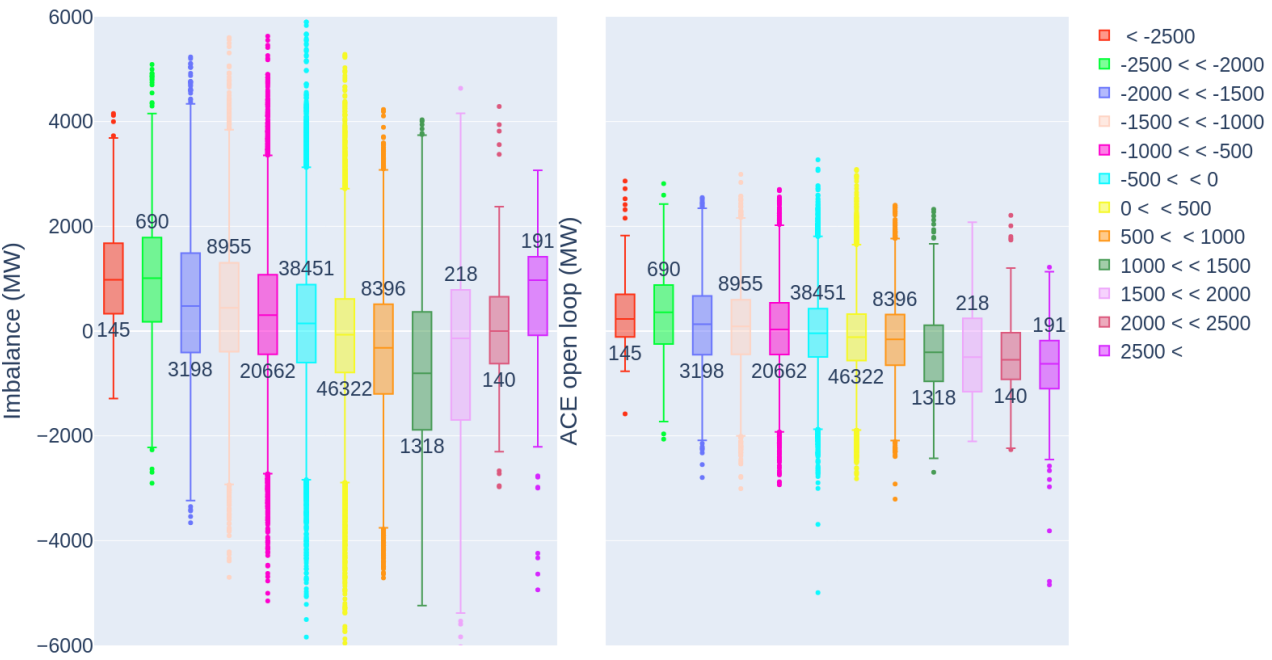}
    \caption{Imbalance (left) and the open-loop ACE (right) by bins of wind power forecast errors.}\label{fig:deseq_ACEol_EO_bin_err_boxplot}
    \end{subfigure}
    \begin{subfigure}{.5\textwidth}
    \centering
    \includegraphics[width=\linewidth]{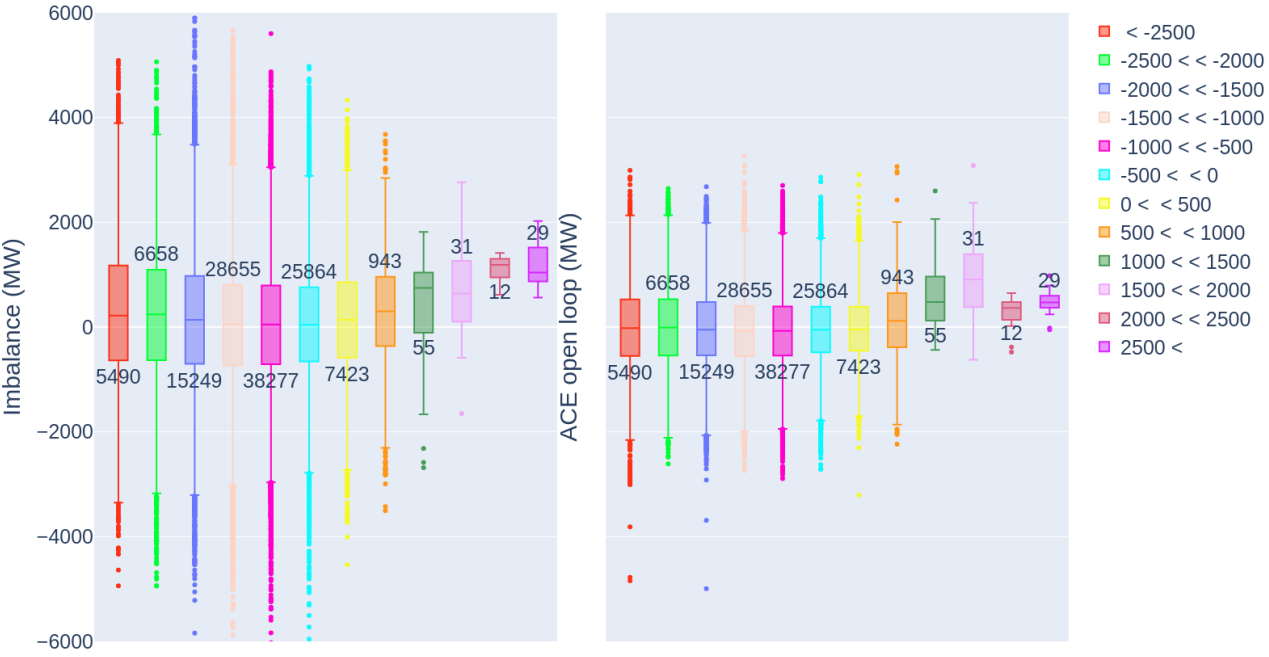}
    \caption{Imbalance (left) and the open-loop ACE (right) by bins of consumption forecast errors.}\label{fig:deseq_ACEol_conso_bin_err_boxplot}
    \end{subfigure}
    \caption{Boxplots of the imbalance (left) and the open-loop ACE (right) depending on the PV (top), wind power (middle), and consumption (bottom) by bins of forecast errors at one hour ahead. The number of values by boxplot considered is displayed. }\label{fig:deseq_ACEol_by_fc_PV_EO_conso_err_boxplot}
\end{figure}

This appendix presents the Figures of the imbalance and open-loop ACE analysis conducted in Sections \ref{sec:imb-analysis-1} and \ref{sec:imb-analysis-2}.
In addition, it also provides the results of the percentiles deviation of Figures \ref{fig:deseq_ACEol_by_fc_PV_EO_boxplot} and \ref{fig:deseq_ACEol_by_fc_PV_EO_conso_err_boxplot}, which is the difference between the 99th and 1st quantiles ($\Delta Q = Q_\text{99} - Q_\text{1}$).

%%%%%%%%%
% Delta_Q PV/EO/conso en FC
%%%%%%%%%%%
% Delta_Q PV FC results 
This paragraph presents the results of $\Delta Q$ related to Figure \ref{fig:deseq_ACEol_by_fc_PV_EO_boxplot} and supporting the conclusions drawn in section \ref{sec:imb-analysis-1}.
The quantile deviation of imbalance ($\Delta Q$) varies slightly across different PV LF values (Figure \ref{fig:deseq_ACEol_by_fc_PV_boxplot}), producing the following values (in MW) [5775.2, 6346.1, 6185.8, 5649.2, 5979.2, 5851.1]. However, this deviation remains nearly the same for LF values less than 0.1 and greater than 0.5. Meanwhile, the $\Delta Q$ of the open-loop ACE stays relatively constant with the following values (in MW): [3385.7, 3216.0, 3269.1, 3258.6, 3303.5, 3298.6].
%
% Delta_Q EO FC results 
The quantile deviation for the imbalance increases by approximately 1100 MW with wind power LF values (Figure \ref{fig:deseq_ACEol_by_fc_EO_boxplot}), with values recorded at [5421.4, 5742.0, 6816.6, 6642.0, 6643.6, 6516.5]. The trend for the open-loop ACE mirrors this, with $\Delta Q$ showing a sharp increase of 700 MW, with respective values of [3077.1, 3140.7, 3455.7, 3581.8, 3805.0, 3773.4].
%
% Delta_Q conso FC results 
The open-loop ACE's quantile difference ($\Delta Q$) shows minimal variation with the normalized consumption (Figure \ref{fig:deseq_ACEol_by_fc_conso_boxplot}), with values (in MW) of [3206.1, 3404.7, 3300.5, 3493.2]. In comparison, the $\Delta Q$ for the quantile imbalance deviation significantly increases by 1010 MW with consumption, yielding [5650.5, 6010.7, 5944.7, 6660.7] values.
% Boxplots du déséq/ACEol vs facteur de charge EO/PV et conso normalisée

%%%%%%%%%
% Delta_Q PV/EO/conso erreur de prévision
%%%%%%%%%%%
% Delta_Q PV FC results 
This paragraph presents the results of $\Delta Q$ related to Figure \ref{fig:deseq_ACEol_by_fc_PV_EO_conso_err_boxplot} and supporting the conclusions drawn in section \ref{sec:imb-analysis-2}.
%
% Quantiles Delta_Q PV erreur de prévision
Concerning the open-loop ACE distribution variation vs. errors in PV forecasts (Figure \ref{fig:deseq_ACEol_PV_bin_err_boxplot}), the 99\% and 1\% percentiles of the open-loop ACE show a slight increase. Furthermore, the difference between these percentiles $\Delta Q$ is gradually rising, increasing by approximately 1200 MW, with recorded values of [741.4, 2340.7, 2725.2, 3301.8, 3304.5, 3483.7, 3205.3, 3377.9, 3345.1, 3539.5, 3667.8, 3795.6]. 
Regarding imbalance, $\Delta Q$ has increased by around 4000 MW between negative and positive errors, with recorded values of [271.0, 3787.6, 4616.0, 5515.1, 5627.3, 6134.4, 5814.4, 6542.4, 6822.9, 7210.4, 7586.4, 7823.9].
%  Delta_Q EO erreur de prévision
Regarding Figure \ref{fig:deseq_ACEol_EO_bin_err_boxplot}, the difference between the 99th and 1st percentiles progressively increases with the wind power forecast errors, yielding a difference of around 1400 MW, represented by the values [3401.2, 3550.3, 3814.5, 3726.9, 3411.3, 3247.3, 3175.8, 3354.6, 3615.1, 3939.0, 4048.6, 4946.9].
Concerning the imbalance, both the 1st and 99th percentiles tend to increase. $\Delta Q$ rises by approximately 2900 MW between negative and positive wind power forecast errors, as indicated by the values [5114.8, 6811.1, 6760.8, 6547.0, 6020.8, 5880.7, 5698.6, 6408.1, 8160.6, 9943.5, 6755.8, 8017.7].
% Delta_Q conso erreur de prévision
Finally, $\Delta Q$ of the open-loop ACE tends to decrease progressively with the consumption forecast error (Figure \ref{fig:deseq_ACEol_conso_bin_err_boxplot}) with the following values (in MW): [3938.0, 3799.9, 3593.4, 3353.1, 3225.5, 3130.4, 3014.9, 3402.1, 2711.3, 3476.6, 1109.0, 967.8]. 
More surprisingly, the imbalance distribution decreases as the consumption forecast error increases. $\Delta Q$ decreases by approximately 5300 MW as the forecast error increases, as indicated by the values [6707.9, 6749.5, 6750.1, 6281.2, 6030.6, 5636.7, 5372.5, 5717.5, 4360.5, 3897.3, 781.0, 1404.6].

%%%%%%%%%%%%%%%%
% Boxplots du déséq/ACEol vs erreur de prév à échéance 1 heure EO/PV et conso normalisée
%%%%%%%%%%%%%%%%%%%%%%%

\subsection{Training methodology and additional results for the forecasting of imbalance/open-loop ACE}\label{appendix:ML}

This appendix \ref{appendix:ML} presents the training and evaluation methodology, the results of the metric values, and the Figures of sections \ref{sec:imb-forecasting} and \ref{sec:ace-forecasting}.

Due to the limited volume of data available, a \textit{k-fold cross-validation} method has been implemented, with k set to 4. This choice allows for a testing set encompassing four months, which represents about 30\% of the entire dataset. For instance, in the first learning/testing pair, the learning set includes data from May 2022 to March 2023, while the testing set covers the period from April 2023 to July 2023.
The second data pair includes a training period from September 2022 to July 2023, with the testing set comprising data from May 2022 to August 2022. Evaluation metrics are calculated for each testing set and then averaged, which allows for an "artificial" increase in the size of the testing set and leads to more robust results. This approach ensures that testing covers the entire dataset.
The hyperparameters for the HGBR model are as follows: learning rate = 0.1 and max iteration = 200. All other hyperparameters remain set to their default values.
%
% Résultats des métriques pour l'imbalance
%%%%%%%%%%%%%%%%%%%%%%%%
%
% Résultats des métriques pour la prédiction de l'ACE open-loop
%%%%%%%%%%%%%%%%%%%%%%%%
\begin{table}[htbp]
\renewcommand{\arraystretch}{1.25}
\begin{center}
\begin{tabular}{lrrr}
\hline  \hline
&  X1+X2+X3 &  X2 & X3  \\ \hline
\multicolumn{4}{l}{\textbf{Imbalance}} \\ \hline
% Imbalance
$\overline{\text{MAE}}_\text{LS}$ & \textbf{\underline{204.3}} & \underline{225.2}& 757.9\\
\hline
$\overline{\text{MAE}}_\text{TS}$ & \textbf{\underline{221.6}} & \underline{234.5}& 913.7\\
\hline
$\overline{\text{RMSE}}_\text{LS}$ & \textbf{\underline{296.1}} & \underline{332.7}& 974.7\\
\hline
$\overline{\text{RMSE}}_\text{TS}$ & \textbf{\underline{324.7}} & \underline{334.6}& 1181.2\\
\hline
$\overline{\text{PL}}_\text{LS}^{\text{Q}_1}$ & \textbf{\underline{10.9}} & \underline{12.5}& 26.0\\
\hline
$\overline{\text{PL}}_\text{TS}^{\text{Q}_1}$ & \textbf{\underline{14.8}} & \underline{15.2}& 37.9\\
\hline
$\overline{\text{PL}}_\text{LS}^{\text{Q}_{50}}$ & \textbf{\underline{103.5}} & \underline{111.3}& 384.4\\
\hline
$\overline{\text{PL}}_\text{TS}^{\text{Q}_{50}}$ & \textbf{\underline{110.6}} & \underline{116.4}& 453.1\\
\hline
$\overline{\text{PL}}_\text{LS}^{\text{Q}_{99}}$ & \textbf{\underline{9.8}} & \underline{11.2}& 26.5\\
\hline
$\overline{\text{PL}}_\text{TS}^{\text{Q}_{99}}$ & \textbf{\underline{13.2}} & \textbf{\underline{13.2}} & 38.5\\ \hline
\multicolumn{4}{l}{\textbf{Open-loop ACE}} \\ \hline
% open-loop ACE
$\overline{\text{MAE}}_\text{LS}$ & \textbf{\underline{146.2}} & \underline{160.7}& 463.9\\
\hline
$\overline{\text{MAE}}_\text{TS}$ & \textbf{\underline{158.7}} & \underline{169.8}& 536.5\\
\hline
$\overline{\text{RMSE}}_\text{LS}$ & \textbf{\underline{195.5}} & \underline{215.2}& 585.9\\
\hline
$\overline{\text{RMSE}}_\text{TS}$ & \textbf{\underline{214.0}} & \underline{228.7}& 678.3\\
\hline
$\overline{\text{PL}}_\text{LS}^{\text{Q}_1}$ & \textbf{\underline{6.3}} & \underline{6.5}& 15.1\\
\hline
$\overline{\text{PL}}_\text{TS}^{\text{Q}_1}$ & \underline{8.4} & \textbf{\underline{8.3}}& 19.0\\
\hline
$\overline{\text{PL}}_\text{LS}^{\text{Q}_{50}}$ & \textbf{\underline{73.5}} & \underline{80.1}& 84.9\\
\hline
$\overline{\text{PL}}_\text{TS}^{\text{Q}_{50}}$ & \textbf{\underline{79.6}} & \underline{233.5}& 266.9\\
\hline
$\overline{\text{PL}}_\text{LS}^{\text{Q}_{99}}$ & \textbf{\underline{5.9}} & \underline{6.3}& 7.5\\
\hline
$\overline{\text{PL}}_\text{TS}^{\text{Q}_{99}}$ & \textbf{\underline{7.5}} & \underline{16.0} & 19.6\\
\hline  \hline
\end{tabular}
\caption{The MAE, RMSE, and PL metrics are averaged over the four splits of the four-fold cross-validation process using an HGBR model to predict the imbalance (top) and the open-loop ACE (bottom). The PL metric presents scores for the 1st, 50th, and 99th percentiles. The scores are provided for the learning (LS) and testing sets (TS). Values that are both underlined and bolded are ranked first, while those that are only underlined are ranked second.
}
\label{tab:imb-ace-results}
\end{center}
\end{table}
%
%%%%%%%%%%%%%%
% resultats prediction déséquilibre timeserie
%%%%%%%%%%%%%%%
%
%%%%%%%%%%%%%%
% resultats prediction déséquilibre scatter plot
%%%%%%%%%%%%%%%

\end{document}